\documentclass[12pt]{article}
\usepackage{amsmath}
\usepackage{amsfonts}
\usepackage{amssymb}
\usepackage{graphicx}

\ifx\pdfoutput\@undefined\usepackage[usenames,dvips]{xcolor}
\else\usepackage[usenames,dvipsnames]{xcolor}
\IfFileExists{pdfcolmk.sty}{\usepackage{pdfcolmk}}{} 
\fi
\usepackage[plainpages=false,pdfpagelabels,pagebackref=false,naturalnames=true,hyperindex=true,pdftitle={What is Computation?},pdfauthor={Hector Zenil}]{hyperref}
\hypersetup{colorlinks=true,
urlcolor=Cerulean,linkcolor=BrickRed,citecolor=RoyalBlue,a4paper,
  pdfpagemode=None,
  pdfstartview=FitH}
\usepackage[all]{hypcap}

\begin{document}

\title{What is Nature-like Computation?\\A Behavioural Approach and a Notion of Programmability\footnote{Philosophy \& Technology, Springer 2012 (special issue on History and Philosophy of Computing)}}
\author{Hector Zenil\footnote{h.zenil@sheffield.ac.uk}%\\Behavioural and Evolutionary Theory Lab\\Dept. of Computer Science, University of Sheffield, UK, and
\\Institut d'Histoire et de Philosophie des Sciences et des \\Techniques (Paris 1/ENS Ulm/CNRS), and\\LABORES for the Digital and Natural Sciences,\\Paris, France.
%h.zenil@sheffield.ac.uk,  hector.zenil-chavez@malix.univ-paris1.fr
}
\date{}

\maketitle

\begin{abstract}
The aim of this paper is to propose an alternative behavioural definition of computation (and of a computer) based simply on whether a system is capable of reacting to the environment---the input---as reflected in a measure of \emph{programmability}. This definition is intended to have relevance beyond the realm of digital computers, particularly vis-\`a-vis natural systems. This will be done by using an extension of a phase transition coefficient previously defined in an attempt to characterise the dynamical behaviour of cellular automata and other systems. The transition coefficient measures the sensitivity of a system to external stimuli, and will be used to define the susceptibility of a system to being (efficiently) programmed.\\

\end{abstract}

\section{The question of computation and computing}

Despite attempts to formulate one spanning several decades, no agreed-upon definition of computation exists to this day \cite{putnam,copeland,shagrir,smithb,smith2,smith3,piccinini}. Legitimate objections have been raised, for instance, to representationalist and functionalist definitions \cite{putnam,searle,fodor,piccinini,piccinini2}. No existing account of computation seems free of the following requirements: that the system's specification be known, that a model representation or mappings among states or functions be available. Which makes it difficult to characterise natural computation (some accounts simply decree that natural computation constitutes computation as such). Here we advance what to our knowledge is a novel approach based on the behaviour of a system, very much in the spirit of Turing's response when asked about the applicability of the notion of intelligence to entities other than humans or animals, in particular to computing machines. Formal thought about computation begins with machines and then proceeds to consider natural processes (such as brains and physical phenomena), which runs contrary to the direction of history, since the original `computers', the entities engaged in computation when Turing began thinking about electronic computers, were actually human beings. Perhaps this paradox has to do with the centrality of the digital in today's  understanding of computation, as against a possibly more general notion of computation covering other than digital.

 The most important notion of computation has been the notion of digital computation, and the most important feature of digital computation is universality. Turing's abstract idea of a universal computer turned out to be technologically feasible, showing that though physics may not compute, it at least supports computation, as we have built devices whose behaviour, despite being governed by the laws of physics, effectively implements general-purpose digital computation. Even though other forms of computation may exist or have been advanced---quantum computation, for instance---they are measured against the digital model, and no consensus exists as regards the notion of universality except with reference to digital computation (analogue computation, for example, has no standard definition of universality. Thus since Turing, the concept of computation has for the most part been couched in the form of the concept of digital mechanical computation. The seminal concept of computational universality can clearly be associated with the amenability of a machine to being (re)programmed, with the existence of a general-purpose machine that can emulate any other machine of the same class and is hence deeply related to the notion of \emph{programmability}. In this paper we place the concept of programmability at the centre of the discussion, with a view to extending the notion of computing to unconventional and natural computation. Programmability here is invoked in a slightly different manner than its standard usage (e.g., a specific--purpose TM is programmable on the present account).

The study of the limits of computation has succeeded in generating insight into what computation might be. The borderline between the decidable and the undecidable has provided an essential intuition in our search for a better understanding of computation. One can, however, wonder just how much can be expected from such an approach, and whether other, alternative approaches to understanding computation may complement the knowledge and intuition it affords, especially vis-a-vis modern uses of the concept of computation in the context of nature and physics, corresponding to situations in which objects or events are seen as computers or computations. 

One such approach involves not the study of systems lying ``beyond'' the uncomputable limit (also called the Turing limit), but rather the study of the minimum requirements for reaching universal computation, through a focus on the `smallest' possible systems capable of universal computation---how easy or complicated it is to build a universal Turing machine, and how efficient such a machine is. This minimalistic bottom-up approach is epitomised by Wolfram's programme \cite{wolfram} together with its interesting older \cite{minsky,yuri,yuri2,watanabe} and more recent incarnations \cite{wolfram,cook,smith,margenstern,neary,baiocchi}. Putnam \cite{putnam} and Searle have advanced strong objections based on the argument that an overly broad notion of computation leads to trivialisation, where for example, every system realises every Turing-computable function (strong pancomputationalism), and where arbitrary state mappings and encodings yield meaningless definitions of computation. 

One can think of formal semantics as an approach to defining computation through programming languages and models of computation, which makes a distinction between syntax and semantics, mapping programs onto mathematical objects describing the relationship between the syntax and the model of computation (thus being model dependent). According to the semantics approach, a computation is a function that maps input onto output \cite{scott}. Syntax defines the correct form for valid programs and semantics determines what (if anything) they compute. In other words:

\begin{center}
Computation $ = $ PL Syntax $+$ PL Semantics
\end{center}

With PL meaning programming language. There are several widely used techniques (e.g. algebraic, axiomatic, denotational, operational, and translational) for the description of the semantics of programming languages, all of which deal with their behaviour. The distinction is often made between syntax, concerned with what constitutes a program, and semantics, concerned with the question of what a program computes, or what an expression means and whether or not two expressions are equivalent.

In most accounts of computational processes as realised by physical mechanisms, it is also often assumed that there is a one-to-one correspondence between causal physical states and the states of a computation defined by some abstract model in which these states can be represented. Putnam \cite{putnam} and Scheutz \cite{scheutz} discuss the issue of the correspondence of abstract/computational state to physical state. 

The traditional mapping-states definition of physical computation is probably inspired by formal semantics, in that it requires that a mapping be established between a model and a physical system, meaning that states and events in the model are used to label states and events observed in the system treated as mathematical objects. For example, if $\phi$ is a Turing universal computing machine, for any other computational process $\omega$, there is an effective mapping $\mathcal{M}$ such that any input $x$ for $\omega$ can be encoded as an input $\mathcal{M}(x)$ for $\phi$, so that after $\phi$ has performed its computation, $\phi ({\mathcal{M}(x)})$ can be decoded to the answer that $\omega$ would have given for $x$, that is $\mathcal{M}^{-1}(\phi(\mathcal{M}(x)))=\omega(x)$ if $\omega$ halts.

However, one can rarely find or justify such mappings $\mathcal{M}$ in natural systems (independently of whether they exist), models of natural systems are often not easily amenable to formal analysis and simulation at different scales. In general it is arbitrary to, for example, map a ``natural state" to a halting state for the standard definition of digital computation to apply, nor can we always know what path nature has taken to produce a given outcome, regardless of whether or not we see this path as constituting a computation (and even independently of whether there is a causal connection).

The syntactic approach to defining computation is closely related to the common (and often informal) view that computation is information processing, which in turn is related to what I think is an informal gesture toward ``programmability". The direct connection between computation and information processing is not trivial at all and Floridi has advocated to make a clear distinction between two variations of information of semantic content \cite{floridi} where the observer plays an important role. Floridi's account of one of two types of information \cite{floridi2}, that is \emph{instructional information}, may be related to what this approach is proposing regarding the existence of "an interpreter that transforms what is in a system into a set of instructions to be executed by a computer, showing that the system can be said to compute. The importance of this comment stems from the role that propagation/processing of information plays in the present account.

From the point of view of this behavioural approach to computation, the syntactic approach falls short of achieving its own objective, viz. distinguishing what computation is from what it is not, because it is not at all clear how one can or cannot find a mapping between the states of any given (even natural or physical) system and a computational one in the broad contexts we are interested in. The semantic approach, however, appears to accommodate some common intuitions about what does and does not count as a computing system, but it requires that relations be established between computational states. For example, it can be made to work by introducing the notion of an interpreter. If an interpreter exists to transform what is in a system into a set of instructions to be executed by a computer, then the system can be said to compute. This, however, is not an easy task when it comes to making such mappings between natural and state-oriented computing systems. 

David Deutsch \cite{deutsch}, has defended the position that computers are physical objects, and computations are physical processes governed by the laws of physics. However, the theory of computation has traditionally been studied almost entirely in the abstract, as a topic in pure mathematics. If computers are closer to physical and natural processes one may expect to face some of the same challenges in the approach to defining computation for abstract systems than for natural systems. Turner \cite{turner}, for example, points out De Millo's ``argument from complexity" \cite{demillo}. That is, ``that the complexity of many contemporary computational systems irrespective of their ontological nature, demands that they be treated as physical systems", because there is no practical approach to verifying the correctness of a sophisticated computer program (for example, an air traffic controller) without treating it as a ``natural system", that is performing experiments and observing their outcomes (and this is in practice one of the pillars of software verification in techniques such as \emph{randomized testing}), and not as a ``mathematical system" for which one can verify its correctness without having to run the program (very much along the lines of Wolfram's Principle of Computational Equivalence \cite{wolfram} that seems to synthesise and generalise this position).

We advance here a metric of behaviour of computation along these lines, distinguishing objects to which some degree of computation can be assigned according to how they behave, particularly how they can be programmed, hence placing programmability at the centre of the discussion and definition of computation. Clearly, by such a definition one may now not call something a computer if for any input it leaves it unchanged, or if for any input the same output is always produced, which is what we take as the contrary of the action of ``being programmed". Between these two cases, however, we think there is room for a behavioural definition. Thus a system will be a computer depending on its capabilities to react to external stimuli related to the notion of ``programmability". The usual feeling with things such as calculators is that one \emph{does not program a calculator} because it is a ``specific-purpose computer" but still a calculator is highly sensitive to external stimuli and depends completely on the input (numbers and operations) to determine the output, hence it is in this sense highly programmable. In our context, programmable does not mean therefore to rewire a system but to make it behave in a desired way (e.g. adding or multiplying given numbers). We think this is the spirit behind the concept of universality, that is, that hardware or software rewiring is at some fundamental level equivalent. The idea of a universal Turing machine is precisely to have a general-purpose machine to be able to behave as any specific-purpose Turing machine. Turing showed that one could encode such a specific-purpose machine as data for the universal Turing machine. In other words, a Turing universal computer has no problem with behaving as an electronic calculator without rewiring the universal computer. This also goes in the other direction. For every specific-purpose program running on a universal Turing machine, one can build a machine implementing the specific-purpose program. Hence no fundamental distinction between hardware and software exists.

One immediate reaction, and a possible objection to a behavioural approach to computation, is the applicability of such a behavioural  and observer-oriented definition to offline computing systems (e.g., batch processing systems). This is a legitimate objection, and in Ref.~\cite{sapere} I discuss this and other objections, which also apply to other behavioural approaches to other notions, such as the notion of intelligence, and the Turing test. The immediate answer to the objection is that even a batch process would have to be programmed at some time, so its ``black box" status, ``hiding" a computing process from the observer, is only transitional. This is discussed further and addressed in Ref.~\cite{sapere} and \cite{context} by way of a definition of limit behaviour.

 A second immediate reaction is whether placing programmability at the centre of a definition of computation involves too strong an assumption, as there are artificial and natural systems that may not look \emph{programmable} which one would nevertheless grant are capable of computing (e.g. discrete neural networks). I think this objection proceeds from a conflation between the standard meaning of programming and the behavioural one I am advancing here. While it is true that for many artificial and natural systems a concept of programmability is difficult to determine, the concept of programmability advanced in this paper amounts to whether one can, by any means, make a system behave differently from the way it was already behaving. In this sense, for example, a logic circuit or a batch process may not qualify as computation if these are unable to react to external stimuli, or if the observer is unable to witness such an interaction if it happened in the design or the launch of a computing process, which brings us back to the previous objection and to the discussion in Ref.~\cite{sapere} and \cite{context}.

\section{A behavioural approach}

Significant effort has been invested in defining computation in denotational, operational and axiomatic terms. For example, while most approaches prove that their definitions of a computation and the universe of operations they are able to compute coincide (leading to the Church-Turing thesis), some have adopted operational approaches \cite{dershowitz} which raise the question of whether their definitions are simply too broad. An axiomatic approach has also been developed, with some interesting results \cite{gandy,sieg}. Nevertheless, some authors have extended the definition of computation to physical objects and physical processes at different levels of physical reality \cite{wheeler,feynman,fredkin,wolfram,deutsch,lloyd} ranging from the digital to the quantum. in Ref.~\cite{wolfram}, for instance, Wolfram states that ``\ldots all processes, whether they are produced by human effort or occur spontaneously in nature, can be viewed as computations", a definition which is vulnerable to the criticisms of Searle \cite{searle} and Putnam \cite{putnam}. 

Klaus Sutner \cite{sutner} has this to say in regards to Wolfram's conception of computation in nature: ``This [Wolfram's] assertion is not particularly controversial, though it does require a somewhat relaxed view of what exactly constitutes a computation---as opposed to an arbitrary physical process such as, say, a waterfall." However, the work of several of the aforementioned physicists and computer scientists does indeed allow us to ask whether natural and physical systems are (or can be viewed as) computational processes. In fact the claim does not have the same content when advanced by, say, Wolfram rather than by Lloyd, as the former calls for digital computation while the latter makes use of quantum computation. 

But to make sense of the term ``computation'' in these contexts, I propose a behavioural notion of nature-like computation but meaningful in broader contexts, independent of representation and possible carriers, and classical in the sense that doesn't require any other model but the traditional digital one even if it doesn't need to commit to any specific model (and that I claim is the most distinguishable property and advantage of this proposal, that is model-specific independent). This will require a measure of the degree of programmability of a system by means of a compressibility index ultimately rooted in the concept of algorithmic complexity.

The behavioural approach takes this abstraction to the limit (keeping it physical as opposed to mathematical), with its central question being whether one can program a system to behave in a desired way. This approach, which bases itself on the extent to which a system can be programmed, tells us to what degree a given system resembles a computer. It can serve as an epistemological framework for interpreting the computational nature of a system in the broader modern sense of computation, particularly in a physical context.

 As suggested by Sutner \cite{sutner}, it is reasonable to require that any definition of computation in the general sense, rather than being a purely logical description (e.g. in terms of recursion theory), should capture some sense of what a physical computation might be. While Sutner's suggestion \cite{sutner} has similar motivations to ours, it differs from ours in that his aim is to map the behaviour of a system to the theory of computation, notably computational degrees. Sutner aligns his approach with his reading of the following claim made by Searle: \cite{searle} ``Computational states are not discovered within the physics, they are assigned to the physics." Sutner adds ``A physical system is not intrinsically a computer, rather it is necessary to interpret certain features of the physical system as representing a computation." This obliges Sutner to take into consideration the act of interpretation of a physical system as well as the observer. Sutner's observer's language maps the physical object to an interpretation of what the object does as a computational process. In Sutner's view the observer may in the process of interpretation slightly modify the computation without adding to or carrying out the computation attributed to the physical object. One can see Sutner's model as consisting of a pair of coupled automata, where one is the physical object and the other the observer. The observer is defined as an automaton constrained in computational power, capable of mapping (interpreting)---by way of a transducer---a physical object onto a computational process using electrical signals. As in my behavioural approach, the observer plays an important role here, one that is often overlooked in traditional approaches to the definition of computation (indeed even ruled out). This is discussed further in Ref.~\cite{sapere}. Our approach is only concerned with the qualitative character of a computational process and not its inner workings. 

Does the question behind computation concern what enables universality in a computational setup and how pervasive it is? As Wolfram \cite{wolfram} has long claimed (and captured in his intuitive Principle of Computational Equivalence), and as Davis has more recently acknowledged \cite{davis}, it takes very little to reach universality. In fact it is now clear that it is more difficult to devise non-trivial systems that are not Turing universal than it is to devise universal ones. We know, for example, that systems that nobody ever designed as computers are able to perform universal computation, for example Wolfram's Rule 110 \cite{wolfram,cook}, and that these, like other remarkably simple systems, are capable of complex behaviour and universal computation (e.g. Conway's Game of Life or Langton's ant). These systems may be said to readily arise physically, as they have not been deliberately designed. 

As suggested by Javier Blanco \cite{blanco}, a program can be interestingly defined as that which turns a general-purpose computer into a special-purpose computer. This is not a strange definition, since in the context of computer science, a computation (and not even only digital but in general) can be typically regarded as the evolution undergone by a system when running a program. However, while interesting in itself, and not without a certain affinity with our approach, this route through the definition of a general-purpose computer is a circuitous one to take to define computation. For it commits one to defining computational universality before one can proceed to define something more basic, which ideally should not depend on such a powerful (and even more difficult-to-define) concept. Universality is without a doubt the most important feature of computation, but every time one attempts to define computation in relation to universal computation, one ends up with a circular statement [computation is (Turing) universal computation], thus merely leading to a version of a CT thesis.

\section{Programmability}

\subsection{Cellular Automata as a case study}

A cellular automaton (CA) is a computational model that has proved to be an interesting object of study, both as a computational device per se and for modelling all kinds of phenomena \cite{wolfram,illachinski}. A CA consists of an array of cells where each takes a value from a finite set of states. Every cell updates its value depending on the state of its neighbouring cells. Hence the global behaviour of the automaton depends on the local interaction of its cells. An Elementary Cellular Automaton (ECA) is a finite automaton defined in a one-dimensional array. The automaton assumes two states, and updates its state in discrete time according to its own state and the state of its two closest neighbours, all cells updating their states synchronously.

As demonstrated by Wolfram \cite{wolfram}, the evolution of a system like a cellular automaton can be viewed as a computation. As shown in Ref.~\cite{wolfram} (page 638), ECA Rule 132 (denoted from now on as R132, R0, R30, etc.) is a simple cellular automaton whose evolution effectively computes the remainder after division of a number by 2. Starting from a row of $n$ black cells, 0 black cells survive if $n$ is even, and 1 black cell survives if $n$ is odd. So in effect this cellular automaton can be viewed as computing whether a given number is even or odd. Wolfram provides other CA examples computing functions in the traditional sense (e.g. R94 as enumerating even numbers; R62 that can be thought of as enumerating numbers that are multiples of 3; the central column of the pattern of R129 that can be thought of as enumerating numbers that are powers of 2; or a CA with 16 states, as capable of computing prime numbers).

 The CA community has developed a strong intuition for determining the ability of a CA to transmit information and be considered a candidate for universal computation. Evident properties of rules like the Game of Life \cite{conway} (a 2-dimensional cellular automaton proven to be computationally universal) and of rules like R110 \cite{wolfram}, (a one-dimensional nearest neighbourhood) simple cellular automata, are structures persisting over time but sensitive to perturbations. These structures transmit information through a system, for example, in the form of characteristic gliders and all sorts of other well-known structures. These structures are unpredictable in a fundamental way if the system in question is capable of universal computation (as we will learn below from the work of G\"odel and Turing). Predictable rules, or rules with no persistent structures, are often dismissed as incapable of carrying messages and behaving as universal computers. Nevertheless, CAs computing in a one-dimensional space, with only 2 states and nearest neighbour, already have sufficient internal richness, despite  this simplicity, to simulate a cyclic tag system for implementing a universal computing device \cite{cook,wolfram}.

Wolfram noticed \cite{wolfram} this richness, and by careful visual inspection of the evolution of two-dimensional space-time orbits, he was able to classify all the various behaviours into 4 general classes for CA starting with a random initial condition. A measure based on the change of the asymptotic direction of the size of the compressed evolutions of CA (but not limited to CA) for different initial configurations (following a proposed Gray-code enumeration for one-dimensional systems) was presented in Ref.~\cite{zenil}. It gauges the resiliency or sensitivity of a system vis-\`a-vis its initial conditions. This measure led to an interesting characterisation and classification of CA, which when applied to ECA, yielded exactly Wolfram's four classes of systems behaviour. The coefficient works approximating the algorithmic complexity (by compression) of the different evolutions through time of systems starting from different initial configurations.

\subsection{The metric}

On the basis of the principles of algorithmic complexity, one can try to characterise the behaviour of the system \cite{zenil} as approximated in equation \ref{index}. If the evolution is too random, for example, a compressed version of the evolution of the system won't be much shorter than the length of the original evolution itself (one may argue that the complexity of the system is the same, but this is not true, as the complexity of the closed system includes the input of the system). As shown in Ref.~\cite{zenil} this characterisation is not only possible but seems to provide interesting information about the systems (phase transition detection, rate of information transmission, sensitivity, etc). A classification based in the phase transition coefficient as defined in Ref.~\cite{zenil} and here in equation \ref{index}, places at the top systems that can be considered to be both efficient information carriers and highly sensitive (hence related to a measure of ``programmability"), given that they react succinctly to input perturbations. Systems that are too perturbable, however, do not show phase transitions and are grouped as inefficient information carriers along with rules displaying only trivial behaviour. The efficiency requirement is to avoid what is known as \emph{Turing tarpits} \cite{perlis}, that is, systems that may be capable of universal computation but are actually very hard to program. This means that there is a difference between what can be achieved in principle and the practical ability of a system to perform a task. This approach is therefore sensitive to the practicalities of programming a system rather than to its potential theoretical capability of being programmed. 

The first notion to advance is the notion of algorithmic complexity (Kolmogorov-Chaitin or program-size complexity), defined as follows\cite{kolmo,chaitin}:

\begin{equation}
\centering
K_T(s) = \min \{|p|, T(p)=s\}
\end{equation}

That is, the length of the shortest program $p$ that outputs the string $s$ running on a universal Turing machine $T$) \cite{kolmo,chaitin}. A technical inconvenience of $K$ as a function taking $s$ to be the length of the shortest program that produces $s$, is its non-computability, proven by reduction to the halting problem. In other words, there is no program which takes a string $s$ as input and produces the integer $K(s)$ as output. This is usually taken to be a major problem, but one would expect a universal measure of complexity to have such a property. The measure was first conceived to define randomness and is today the accepted objective mathematical measure of complexity, among other reasons because it has been proven to be mathematically robust (in that it represents the convergence of several independent definitions). The mathematical theory of randomness has proven that properties of random objects can be captured by non-computable measures. One can, for example, approach $K$ using lossless compression algorithms that detect regularities in order to compress data. The value of the compressibility method is that the compression of a string as an approximation to $K$ is a sufficient test of non-randomness. If the shortest program producing $s$ is larger than $|s|$ the length of $s$, then $s$ is considered to be random. 

Let $C$ be the approximation to $K$ (given that $K$ is non-computable) by any means, for example, by using lossless compression algorithms. Let's define the function $f$ as the variability of a system $M$ as the result of fitting a curve $\phi$ (by regression analysis) to the data points produced by different runs of increasing time $t^\prime$ (for fixed $n$) up to a given time $t$, of the sums of the differences in length of the approximations to Kolmogorov complexity ($C$) of a system $M$ for inputs $i_j$, $j\in\{1, \ldots, n\}$, divided by $t(n-1)$ (for the sole purpose of \emph{normalising} the measure by the system's ``volume," so that one can roughly compare different systems for different $n$ and different $t$). Formally,

\begin{equation}
\centering
f(M,t,n) = \phi\left(\frac{\sum_{j=0}^{n-1} |C(M_t(i_j)) - C(M_t(i_{j+1}))|}{t(n-1)}\right)
\end{equation}

Where $M_t(i)$ is a system $M$ running for time $t$ and initial input configuration $i$. For one-dimensional input binary systems, a natural numbering scheme devised in Ref.~\cite{zenil} based on the Gray-code is an example for one-dimensional systems. 

At the limit $\mathbb{C}$ captures the behaviour of $M_t$ for $t \rightarrow \infty$, but the value of $\mathbb{C}_n^t$ depends on the choices of $t$ and $n$ (we may sometimes refer to $\mathbb{C}$ as assuming a certain $t$ and $n$), so one can only aim to capture some average or asymptotic behaviour, if any (because no convergence is guaranteed). $\mathbb{C}$ is, however, an indicator of the degree of programmability of a system $M$ relative to its external stimuli (input $i$). The larger the derivative, the greater the variation in $M$ and hence in the possibility of programming $M$ to perform a task or transmit information at a rate captured by $\mathbb{C}$ itself (that is, whether for a small set of initial configurations $M$ produces a single significant change or does so incrementally).

Now we can use $f$ to define a system's programmability (first basic definition) measured by the partial derivative with respect to time:

\begin{equation}
\label{index}
\centering
\mathbb{C}_t^n(M)= \frac{\partial f(M,t,n)}{\partial t}
\end{equation}

For example, according to this coefficient $\mathbb{C}$, ECA with rule numbers 0 and 30 are close to each other because they remain the same despite the change of initial conditions (despite the choice of $t$ and $n$), and they are hardly perturbable. The measure indicates that rules like rule 0 or rule 30 (denoted from now on as R0, R30, etc.) are incapable of transmitting information, given that they do not react to changes in the input. In this sense they are alike because there is no change in the qualitative behaviour of these CA when fed with different inputs, regardless of how different the inputs may be---and this is what $\mathbb{C}$ measures. R0, for example, remains entirely blank, while R30 remains mostly random-looking, with no apparent emergent coherent propagating structures (other than the regular and linear pattern on one of the sides). 

On the other hand, rules such as R122 and R89 have $\mathbb{C}$ close to each other because they are sensitive to initial conditions. As shown in Ref.~\cite{zenil}, they are both highly sensitive to initial conditions and present phase transitions which dramatically change their qualitative behaviour when starting from different initial configurations. This means that rules like R122 and R89 can be used to transmit information through a system, from the input to the output.  

Values of $\mathbb{C}$ for the subclass of CA referred to as ECA (the simplest one-dimensional closest neighbourhood) were calculated and published in Ref.~\cite{zenil}, and a further investigation of the relation between this transition coefficient and the computational capabilities of certain known (Turing) universal machines has been undertaken in Ref.~\cite{zeniluniversalca}. We will refrain from exact evaluations of $\mathbb{C}$ to avoid distracting the reader with numerical approximations that may detract from our particular goal in this paper. The aim here is to propose a behavioural definition of computation based on this measure rather than to evaluate specific values that have already been calculated in Ref.~\cite{zeniluniversalca}. 

This transition coefficient will be used to dynamically define computation based on the \emph{degree of programmability} of a system. The advantage of using the transition coefficient $\mathbb{C}$ is that it is indifferent to the internal states, formalism or architecture of a computer or computing model; it doesn't even specify whether a machine has to be quantum, digital or analogue, or what its maximal computational power should be. It is only based on the behaviour of the system in question. It allows us to minimally characterise the concept of computation on the basis of behaviour alone. And in doing so, it allows us to gauge the efficiency of the reaction to external stimuli and the transfer of information by noting the rate at which $\mathbb{C}$ changes. in Ref.~\cite{sapere} we discuss this ``efficiency" property of $\mathbb{C}$ in more detail.

\begin{figure}[htdp]
\label{example1}
\centering
   \scalebox{.35}{\includegraphics{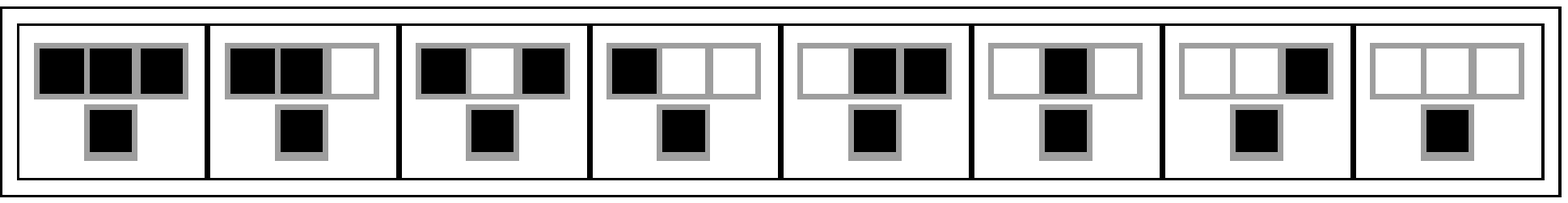}}\\
   \scalebox{.4}{\includegraphics{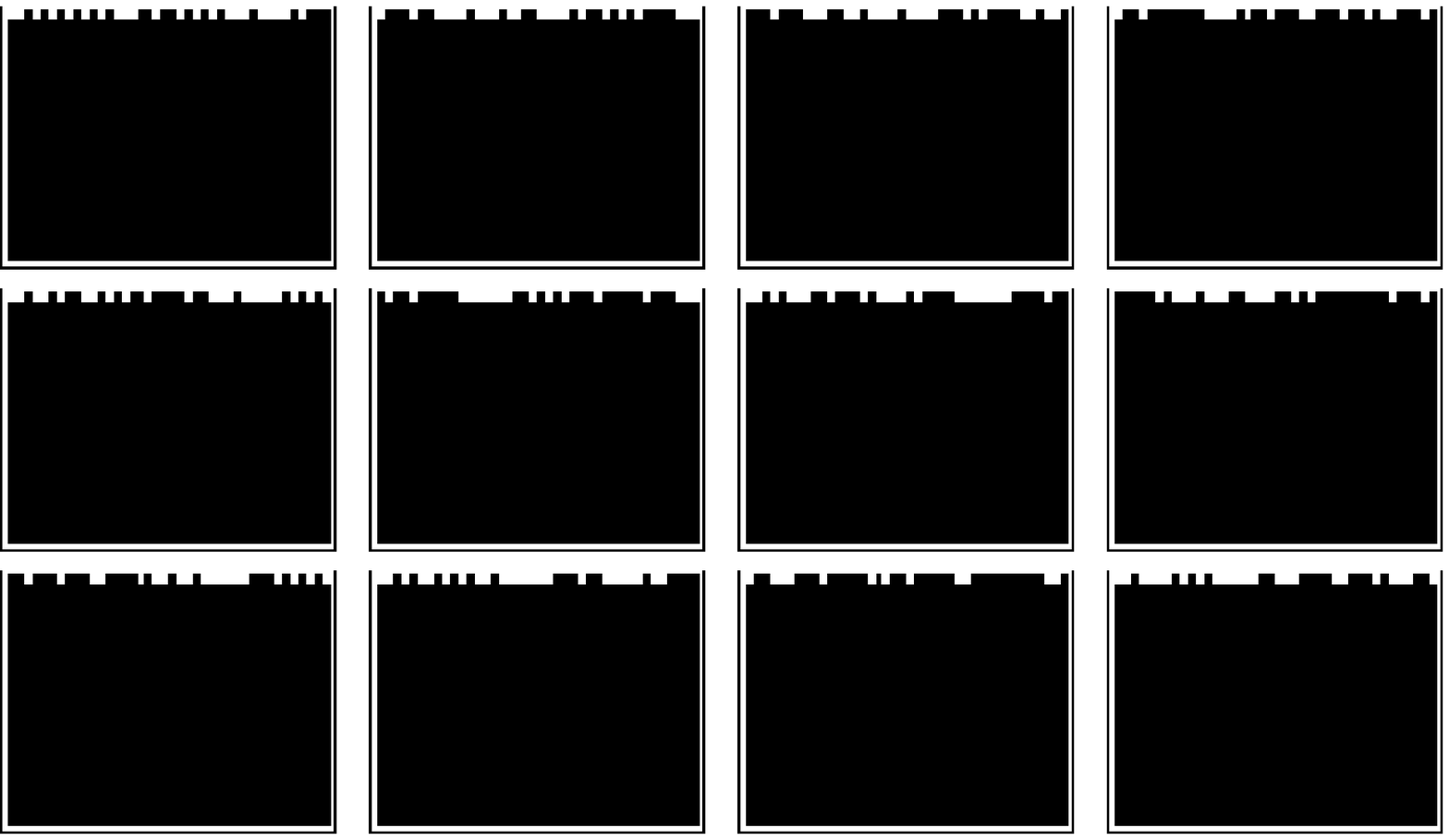}}
\caption{ECA R255 (equivalent by colour inversion to R0; R255 is used here for visual convenience) is stuck, unable to perform any computation---it does not react to any external stimulus. This is an illustration of a $\mathbb{C}$-computer for $\mathbb{C}$ close (or equal) to zero \cite{zenil}. The picture shows a series of evolutions for 12 random inputs, with the cellular automaton rule depicted at the top.}
\end{figure}

\begin{figure}[htdp]
\label{r4}
\centering
\scalebox{.35}{\includegraphics{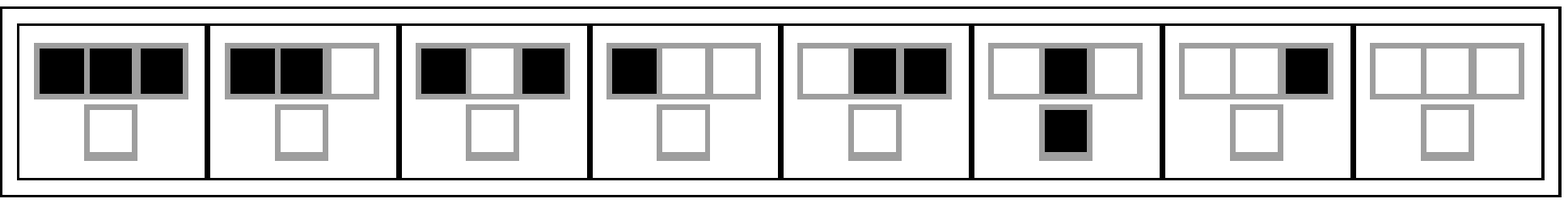}}\\
   \scalebox{.4}{\includegraphics{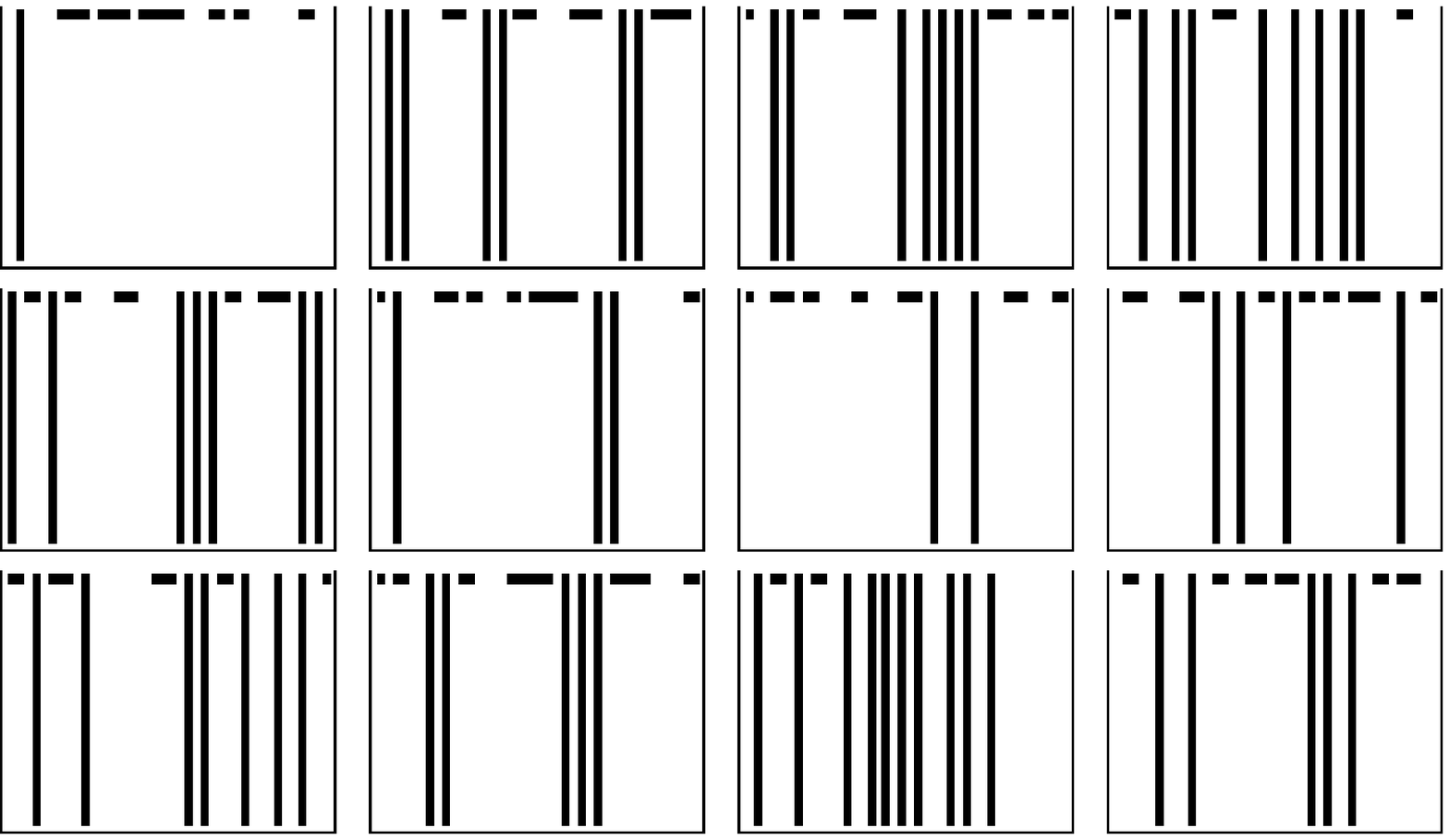}}
\caption{ECA R4 is a kind of program filter that only transfers bits in isolation (i.e. when its neighbours are both white). It is clear that one can perform some very simple computations with this automaton. However, one could not, for example, implement a typical logic gate based on its particular behaviour. It clearly cannot carry out (Turing) universal computation.}
\end{figure}

Let's denote as a $\mathbb{C}$-computer a system with programmability coefficient $\mathbb{C}$ capturing the capability of the system to transfer information from its input towards its output. Under this notation, ECA R255 (Fig. \ref{example1}) is a 0-computer, that is, a computer unable to carry out any operation because it cannot transfer any information from the input to the output (another way to say this is that R255 does not compute), others may compute even if it can be proven to only compute a small subset of the Turing computable functions (see e.g. \ref{r4}). ECA R255 cannot by any means be programmed to perform any task, despite the input. This allows us to answer (in the negative) Chalmers' challenging question \cite{chalmers} prompted by Putnam's objection:  ``Does a Rock Implement Every Finite-State Automaton?". It clearly doesn't under this definition and is in clear contradiction with claims that ``objects compute themselves" \cite{lloyd2} (an objection having to do with scale will be addressed later). The sense of what is required if something is not to be a computer can be captured with the following definition:\\

\textbf{Definition 1.} A 0-computer is not a computer in any intuitive sense because it is not capable of carrying out any calculation.\\

It may be misleading to call a system that does not compute a 0-computer, but it is crucial to this approach to convey the way in which a system is ruled not to be a computer, viz. because its coefficient $\mathbb{C}=0$, the main point of this paper being to distinguish what computation is from what it is not, by means of this alternative ``behavioural" definition.

\begin{figure}[htdp]
\label{110}
\centering
   \scalebox{.35}{\includegraphics{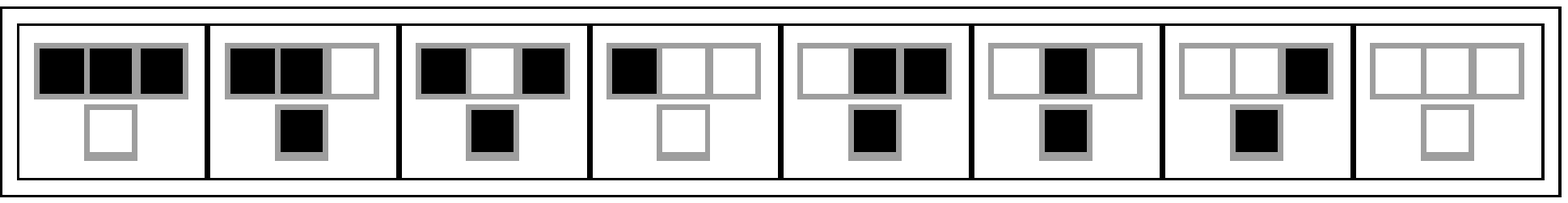}}\\
      \scalebox{.4}{\includegraphics{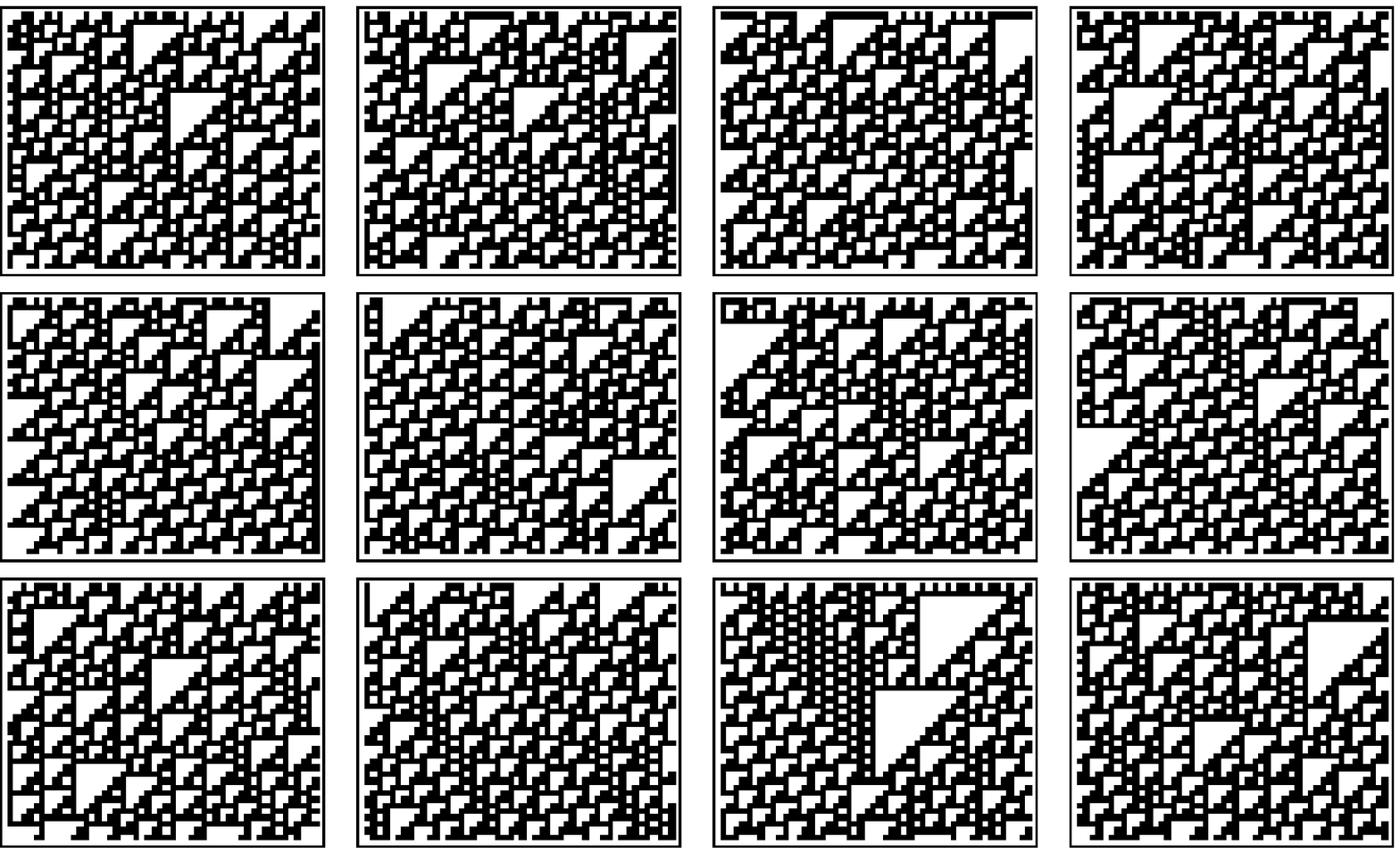}}
\caption{ECA R110 is efficient at carrying information through persistent local structures to the output, reacting to external stimuli. Its $\mathbb{C}_n^t$ value for sensible choices of $t$ and $n$ \cite{zenil} is compatible with the fact that it has been proven that R110 is capable of universal computation for a particular semi-periodic initial configuration \cite{wolfram,cook}.}
\end{figure}

A system capable of (Turing) universal computation (see Fig. \ref{110}) would therefore have a non-zero $\mathbb{C}$ limit value. $\mathbb{C}$ also captures some of the universal computational efficiency of the computer in that it has the advantage of capturing not only whether it is capable of reacting to the input and transferring information through its evolution, but also the rate at which it does so. So $\mathbb{C}$ is an index of both capability in principle and ability in practice. A non-zero $\mathbb{C}$ means that there is a way to codify a program to make the system behave (efficiently) in one fashion or another, i.e. to be programmable. Something that is not programmable cannot therefore be taken to be a computer. 

One can also see that things that seemed to behave like computers but were not called computers can indeed be considered computers under this approach. Mathematical functions, for example, can be considered $\mathbb{C}$-computers for some $\mathbb{C}$ determined by the domain of the function. That a function can be considered a computer does not controvert the theory wherein a computer is defined in terms of a function and a domain, and a function in terms of an algorithm having the input as its arguments and the output as its function evaluation. The calculation of a function, however, seems to require a carrier. Usually that carrier is a piece of paper and a pencil wielded by a human being, but it can also be a physical computer. Can the simple description of the function be considered a computer or a $\mathbb{C}$-computer? I think it should not be. Something static shouldn't be considered to have a behaviour, and I think this can be captured by $\mathbb{C}$. To evaluate $\mathbb{C}$ one needs to actually run a program, otherwise it remains unevaluated (whether it is visible to the observer is the legitimate objection we already mentioned in Section 1, and further discussed in Ref.~\cite{sapere}). Can we not (in principle) think of computations that don't calculate functions? Yes, and this is what this behavioural approach to computation is about. There is no need of representation or even definition of the objects of computation; it is how a system seems to behave which leads us to attribute to it some degree of computation.

This makes for a clear distinction between, for example, a vision of the universe as a mathematical structure and a vision of the universe as a computer. While the latter may account for the physical carrier, implying that the computation is being carried out by the universe itself, it does not seem clear how a mathematical structure can come equipped with the carrier on which it should be executed, unless it becomes a computer program and therefore a computer.

Another example using a 2-dimensional cellular automaton is given in Ref.~\cite{zenilJETAI}, showing that Conway's Game of Life (GoL) indeed has a large enough $\mathbb{C}$ value, which is in agreement with the idea that $\mathbb{C}$ captures the programmability of a system (knowing as we do that GoL is capable of Turing universal computation).

\subsection{Reversibility, 0-computers and conservation laws}

in Ref.~\cite{margolus}, Margolus asserts that reversible cellular automata can actually be used as computer models embodying discrete analogues of classical notions in physics, such as space, time, locality and microscopic reversibility. He suggests that one way to show that a given rule can exhibit complicated behaviour (and eventually universality) is to show (as has been done with the Game of Life \cite{conway} and R110 \cite{cook,wolfram}) that ``in the corresponding `world' it is possible to have computers" starting these automata with the appropriate initial states, with digits acting as signals moving about and interacting with each other to, for example, implement a logical gate for digital computation. Wolfram reinforces this vision by suggesting, through his Principle of Computational Equivalence, that it is indeed the case that non-trivial behaviours inevitably lead to universal computation.

 This does not mean that a system must necessarily be bijective (hence reversible) in its input/output mapping in order to be universal. But it is actually reversible CA with high entropy (number of possible states) which will tend to show the greatest behavioural richness and therefore be considered the best candidates for being classified as computers. In other words, the greater the richness a system is capable of, the greater the $\mathbb{C}$ coefficient it will have. A reversible CA (RCA) has the property that starting it from a random state is like starting from a maximum entropy state in a thermodynamical system, because the RCA is not allowed to get simpler in its evolution, the only way to get simpler being to collapse the number of states, making it irreversible. Entropy in a randomly initiated RCA can only increase, but if it reaches maximum entropy it can't get any more complicated, and so nothing much happens. This is also captured by $\mathbb{C}$, in that the RCA always look the same and are immune to evolutionary changes, presenting homogeneous local entropy everywhere. 

RCA are interesting because they allow information to propagate, and in some sense they can be thought of as perfect computers, indeed in exactly the sense that matters to us. If one starts an RCA from a non-uniformly random initial state, the RCA evolves, but because it cannot get simpler than its initial condition (for the same reason given for the random state) it can only get more complicated, producing a computational history that is reversible and can only lead to an increase in entropy. The RCA, however, is only reshaping the message that it got at the beginning in the form of an initial configuration, and so the amount of information in the RCA evolution remains the same. Which makes it a perfect example of a system with increasing entropy but consistent complexity over time. The algorithmic complexity of the RCA is the same because one can track the RCA back to the original information content represented by its initial configuration. So the state of the CA at any time always carries the same information content. In non-reversible CA, however, information can be lost, and even though the algorithmic complexity of the evolution of a CA is always the same, one cannot recover it a posteriori from any later state. In reversible CA, entropy, like information content, may increase or decrease over time. As Margolus himself states, it is one thing to know that a gas was in one corner at a given state, and another to return the gas from its expanded condition to its original position. It may thus seem that RCA in Wolfram's class III may all be chaotic, but Wolfram \cite{wolfram1984} offers examples of one-dimensional reversible cellular automata exhibiting three types of behaviour of local structures as they propagate in space. 

In nature-like computation, conservation laws are important because the physical carrier on which a computation will be performed is governed by physical conservation laws (laws that conserve physical invariants such as mass, energy, momentum, etc.). In RCA, there are cases where the simplest locally-computable invariants are cells whose values never change, and which are analogous to nature-like conservation laws. That is, laws such that for any given property, the physical state of the system does not change as the system evolves. The simplest RCA capable of doing this are those that ignore their neighbouring cells and only look to the central one, reproducing it identically. One may have doubts about calling these computers because there is no transformation of information whatsoever, with the system just letting pass through it anything that it is fed. Even worse, there are systems that may look as if they are computing the identity function while in fact performing a series of intermediate transformations which lead to the same output a few steps later (again prompting the objection that their performance is relative to the observer). From the behavioural perspective based on the transition coefficient, under the qualitative definition the two would be behaving differently if they delivered their \emph{richness} at different rates, even if they produced the same output. This discussion helps us to see how close these computational systems are to physical phenomena and to purely behavioural descriptions, but also to address some potential concerns raised by the qualitative approach proposed herein. If things could be worse for information processing, given that classical mechanics prescribes determinism in the macroscopic universe, one can extend these worries to the entire world, but this is a subject for a different, though related, discussion.

\section{Behavioural equivalence}

We can then define a system performing computation simply based on its behaviour, as follows:\\

\textbf{Definition 2.} A system $U$ computes if $\mathbb{C}_n^t(U)>0$ for some $t,n>0$\\

Meaning that $U$ can be programmed. Whether $U$ can compute only a subset of computable functions  or all of them will depend not only on $\mathbb{C}$ but also on the details of $U$ that escape the behavioural definition. Yet it is clear that if $U$ is Turing universal, $\mathbb{C}_n^t(U)>0$. This definition accords with a much broader sense of nature and physics-like computation as used in, for example, modern models of physics (to mention but a few examples \cite{fredkin,wheeler,wolfram,deutsch,lloyd}) and natural computation. One can see that there are systems that are not computers under this definition---simple ones are R0 and R255 ECA (see Fig. \ref{example1}). As we know, the equivalence of computations is ultimately undecidable. Even in practice it can only be approached and partially answered, given that the transition coefficient on which the qualitative definition of computation is based is limited by finite resources (reflected by the finite values of $t$ and $n$), providing only an approximate indication of the behavioural programmability of a system, and for $t$ possibly an asymptotic behaviour (no convergence is guaranteed though).

 If two systems have about the same $\mathbb{C}_n^t$ for fixed $n$ and $t$ however, it means that they  react to changes at about the same rate, so it may not only capture the property of transferring information but if information is transferred, it captures the rate at which it does so. Hence by varying $n$ and $t$, one can also possibly soundly define rates of convergence of $\mathbb{C}$. Some of this is also discussed in Ref.~\cite{sapere}.

 Clearly, under this definition behaviour space is less dense than algorithm and program space, because there may be different programs implementing different algorithms but generating the same behaviour. So one can only define two behaviourally equivalent systems as follows:\\

\textbf{Definition 3.} A system $U$ and $U^\prime$ are computationally equivalent in behavioural terms if $\mathbb{C}_t^n(U)=\mathbb{C}_t^n(U^\prime)$ for any $t$ and $n$.\\

Simple examples of a behavioural computational class are $\mathbb{C}$-computers for $\mathbb{C}=0$, i.e. they cannot be programmed, and are behaviourally equivalent. Under Def. 1 and 2, systems that are identified as $0$-computers do not compute, as they are not capable of being programmed. 

Experience tells us that something that behaves in a certain way will continue doing so, and we have empirically established as much in Ref.~\cite{zenilminds}. This can be justified by algorithmic probability, because the longer we observe a computing system, the smaller the chance that its behaviour will change radically. So even though one cannot guarantee a behaviour ad infinitum, algorithmic probability may provide the stability required to make reliable generalisations. Thus one can arrive at a weak Def. 4 by allowing $\mathbb{C}(U)$ to be close enough to $\mathbb{C}(U^\prime)$ as follows:\\

\textbf{Definition 4.} A system $U$ and $U^\prime$ are $c$-computationally equivalent if $|\mathbb{C}(U)-\mathbb{C}(U^\prime)|<c$.\\

It is worth stressing that two systems (or computers) are not the same in any other sense if they have the same coefficient $\mathbb{C}$. $\mathbb{C}$ is a measure of sensitivity (which I understand as the amenability of a system to being programmed); it cannot on its own indicate whether two computers compute the same function, and is therefore a different measure than those available from traditional computability and formal semantics. It can tell when two computers diverge in their behaviour, because for two computers to be the same, a necessary but not sufficient condition is that they must both have the same transition coefficient (or differ by a desired $\mathbb{C}$), which would mean that they have the same capability of reacting to external stimuli, and transmit information at about the same rate. Because $\mathbb{C}$ itself depends on two parameters ($n$ and $t$), this also means that $\mathbb{C}$ can only make comparisons between two systems for fixed $t$ and $n$ (the same runtime and the same number of input configurations). So two $\mathbb{C}$-computers are behaviourally equivalent if they have the same $\mathbb{C}$. 

For the same reason that one cannot tell whether a machine will halt for a given input, one cannot decide whether two computers compute the same function, but one can relate nature-like computation and abstract computation by means of Turing machines as follows: for every $\mathbb{C}$-computer $U$, there exists a program $P$ behaviourally equivalent to $U$, that is, with transition coefficient $\mathbb{C}(U)=\mathbb{C}(P)$ independent of $n$ and $t$, because there exists a universal Turing machine $T$ capable of reproducing the exact behaviour of $U$. 

It is also worth noting that this behavioural definition is cumulative (but not additive), in the sense that a $\mathbb{C}$-computer can be embedded in the workings of another $\mathbb{C}^\prime$-computer for $\mathbb{C}\neq \mathbb{C}^\prime$ (such as in the server room example). If the $\mathbb{C}^\prime$-computer does not impose any behavioural restriction on the $\mathbb{C}$-computer, then clearly $\mathbb{C}^\prime \geq \mathbb{C}$, given that the new computer will be capable of at least $\mathbb{C}$-computation. This is the sense in which one may see R255 as a program in the context of a $\mathbb{C}$-computer with $\mathbb{C} \neq 0$ capable of running R255. If the $\mathbb{C}$-computer is, for example, a universal computer, R255 would be a program but cannot by itself be a computer. The $\mathbb{C}$-computers behaviourally equivalent to R255 would then be all those for which $\mathbb{C}=0$.

\section{What kind of $\mathbb{C}$-computer might the universe be?}

The question of whether a server room containing many racks of servers is in itself a computer in any interesting sense seems to depend on the observer's role. If one wished to apply $\mathbb{C}$ to a server room, it would be legitimate to do so and to consider the room as the set of the $\mathbb{C}$-computers that it contains. Whether doing so is useful or not is another question, but I think that it is not only useful but that it is also common to consider server rooms as assets that can  be commercialised as computers by themselves (computing black boxes for the final users) without getting into the details of the contents of the room and selling its added computing power (this is not very different to farm computing or even current approaches to computation such as multicore systems).

Unconventional computers may also be considered $\mathbb{C}$-computers. When one turns on a lamp the lamp is programmed to do something, in this case to turn on. Even if trivial, it reacts to the input by producing light as the outcome. One can see how the \emph{programmability} notion here is slightly different to the traditional concept because a lamp is a specific-purpose device which cannot be reprogrammed from the traditional point of view, but from this point of view the input changes the behaviour of the system and hence has a programmability degree, even if very limited. But even if the lamp can be considered to react to external stimuli the space of its initial configurations is finite and small (only two possible initial configurations), hence the slope of the differences of the behavioural evolution in time and therefore its $\mathbb{C}_t^n$ is very close to 0 for any $n$ and $t$. If one wants to rule out lamps or fridges as computing devices one would only need to define a threshold for which beyond that threshold a system can be said to compute, while under the threshold it would be discarded. A fridge can be seen as cooling objects that are introduced into it, the output being the cooling---after an interval---of the objects in question. That both a lamp and a fridge can be viewed as $\mathbb{C}$-computers with small $\mathbb{C}$, given that they have limited programmability (to perform a single, specific task), should not be surprising, at least not in light of the definition of a $\mathbb{C}$-computer, nor should it deprive the notion of computation of meaning, as it has been the purpose of this paper to offer a grading system for computation precisely in order to provide meaning to such claims, with the advantage that one can now ask whether a lamp or a fridge is or isn't a computer without trivialising either the question or the answer. Under the behavioural definition advanced herein, they are very limited computing systems only if one wants to keep a threshold of computing very low, as long as it is stated that they are limited in scope by the objective value of $\mathbb{C}$. 

\begin{figure}[htdp]
\label{rule30}
\centering
\scalebox{.35}{\includegraphics{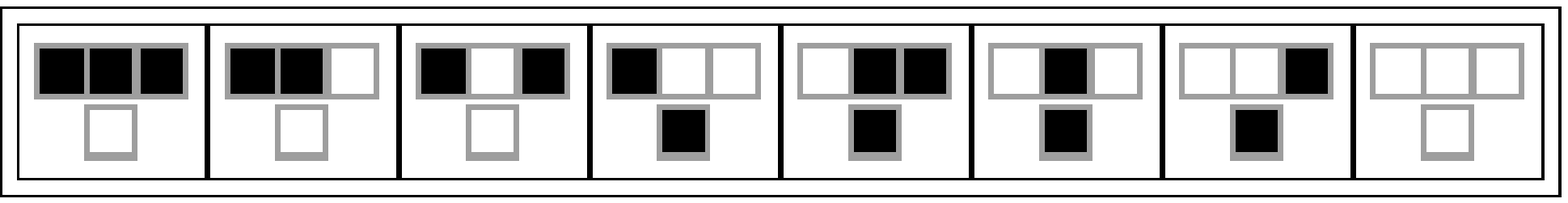}}\\
   \scalebox{.4}{\includegraphics{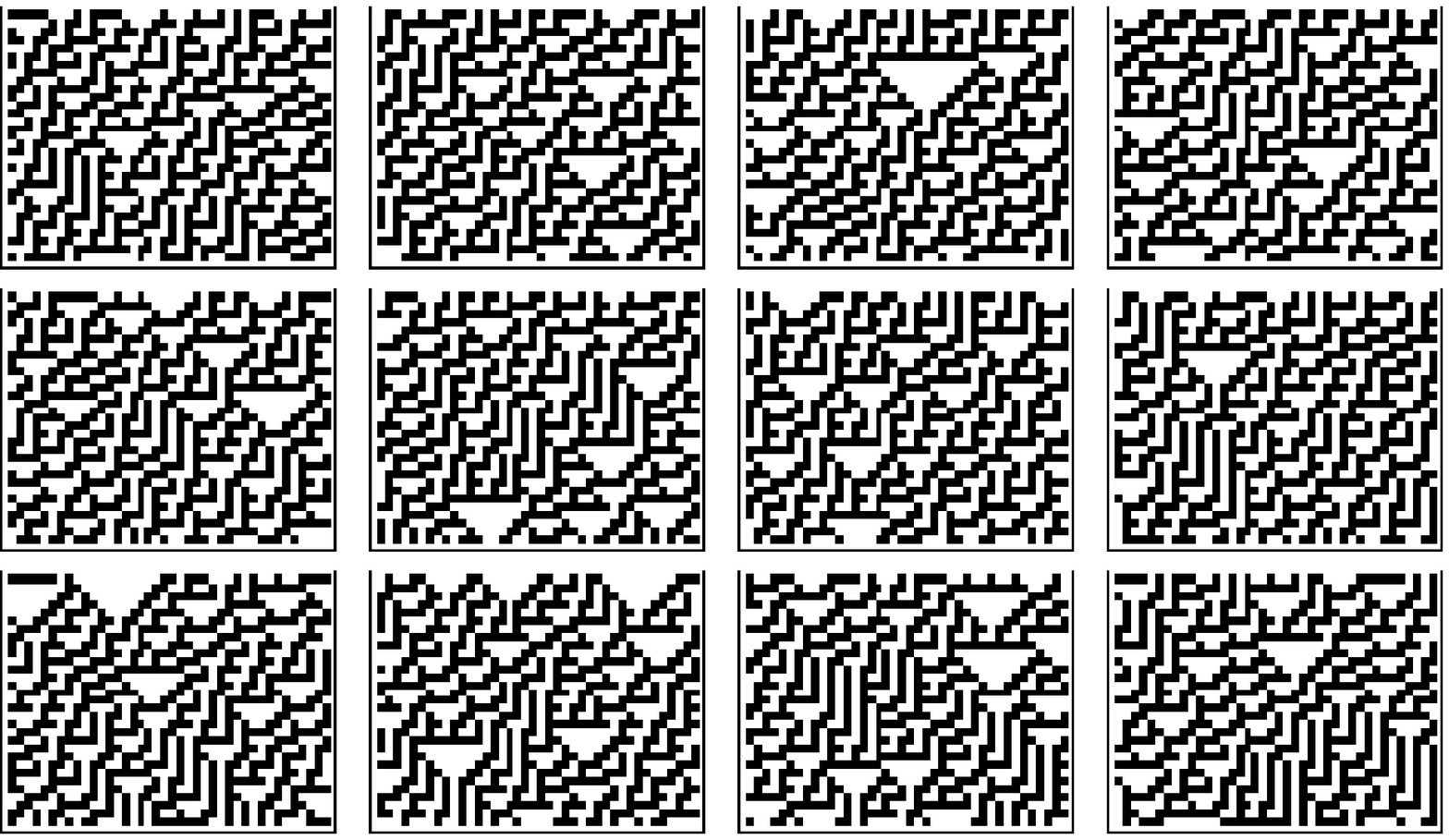}}
\caption{Despite the simplicity of the description of ECA R30 the behaviour of R30 is always random-looking (albeit the leftmost evolution which shows some regularities) even for simple and structured initial conditions. The overall qualitative behaviour of R30 remains unchanged disregarding the initial condition. An open question is whether this rule is ``too hot" to be programmed and used to compute (see \cite{genaro}).}
\end{figure}

One can think of the laws of physics, for example of gravitation, as \emph{carrying out} some sort of \emph{computation}, with the degree of programmability (we are not discussing here whether the model for the physical law corresponds to the real-world, which is a different matter) of such a system limited to performing a particular task, in the case of gravitation  pulling objects toward each other and keeping them in their gravitational trajectory. Classical mechanics guarantees that the system is deterministic, even if that doesn't mean one can predict its workings given any specific parameters (e.g. 3 bodies). There is no fundamental reason, however, for following the approach described herein when assessing whether a system can compute based on its degree of programmability. Still, the fact that one can coarse grain what computation may mean by way of the parameter $\mathbb{C}$, and guarantee that there are both systems with maximal $\mathbb{C}$, and $\mathbb{C}=0$ for systems that can be programmed to do something, and others that cannot be programmed at all and show no reaction to any external stimulus (e.g. see Fig. \ref{example1}), imbues this approach and its definition of computers and computation with sense, particularly in the context of nature-like computation as proposed by some of the aforementioned authors. There are also $\mathbb{C}$-computers for small values of $\mathbb{C}$, meaning that the system can hardly be programmed because it does not transfer information efficiently enough (this may be the case, for example, with R30, see Fig. \ref{rule30}).

This is also related to the recurrent question of whether the universe can be said to compute. In some sense it does, for we know there are $\mathbb{C}$-computers in it capable of universal computation, but we don't really know whether the universe (e.g. as represented by its physical laws) constrains $\mathbb{C}$, a limit broad enough to encompass every possible $\mathbb{C}$-computer for a maximal $\mathbb{C}$ contained in the physical universe. The universe as a whole can be seen and treated in this context as a computer, as it is a $\mathbb{C}$-computer for maximal $\mathbb{C}$, given that it contains all possible $\mathbb{C}$-computers. There is, however, a legitimate strong objection to this view which is discussed in Ref.~\cite{sapere}, given that it is difficult to apply the behavioural measure to a system for which ``external stimuli" is not well defined, vis-\`a-vis the universe, without falling into a contradiction, because any ``external stimuli" is part of the universe. Even if the objection holds, we are not particularly interested in addressing the question of whether the universe is a computer. However, it may be only noted that this objection is again related to the question of the role of the observer and its place.

\subsection{The question of scale}

So far, the object of this behavioural approach to computation has been to provide a reasonable framework for assertions connecting the notion of computation to nature, and how nature may or may not compute, in light of current uses of the term `compute'. Lloyd \cite{lloyd}, for example, claims that since the universe is computing itself, things in the universe would therefore also be computing themselves. Think of the example of a still physical object (e.g. a desk or a sheet of paper). These objects would hardly compute anything at their macroscopic level, say an addition between any 2 numbers, yet they may be constituted at a molecular or atomic scale of particles capable of carrying out all sorts of computations, which unlike the objects, may be programmed, either as part of another system or in themselves. It is clear then that the span of behaviour at that scale is greater than at the scale of the object itself. But does it make sense to say that something computes itself? \cite{lloyd}. It may or it may not. 

In the real world, things are constituted by smaller elements unless they are elementary particles. One therefore has to study the behaviour of a system at a given scale and not at all possible scales, otherwise the question becomes meaningless, as elements of a physical object are molecules, and ultimately atoms and particles that have their own behaviour, about which too the question about computation can be asked. This means that a $\mathbb{C}$-computer may have a low or null $\mathbb{C}$ at some scale but contain $\mathbb{C}^\prime$-computers with $\mathbb{C}^\prime > \mathbb{C}$ at another scale (for which the original object is no longer the same as a whole). A setup in which $\mathbb{C}^\prime \leq \mathbb{C}$ is actually common at some scale for any computational device. For example, a digital computer is made of simpler components, each of which at some macroscopic level but independently of the interconnected computer is of lower behavioural richness and may qualify for a $\mathbb{C}$ of lower value. In other words, the behavioural definition is not additive in the sense that a $\mathbb{C}$-computer can contain or be contained in another $\mathbb{C}^\prime$-computer such that $\mathbb{C} \neq \mathbb{C}^\prime$.

Can R255, for example, be thought of as computing itself as it evolves? Under the qualitative definition, even if R255 is computing itself it cannot be programmed, and so is a 0-computer under our approach, a computer not capable of computation and therefore hardly a computer at all. On the other hand, R255 does not present any problem of scale as it represents itself at all scales. A table, however, is made of smaller components to which may be assigned some specific task, and one may even consider reprogramming the matter of which it is made, in the manner epitomised in the subfield of programmable matter. In which case one may say that the table is computing itself, since it could be computing something else out of its atoms. So the definition of a $\mathbb{C}$-computer is scale-dependent and its implementation in the real world is subtle, yet at the abstract level it seems to correspond to an interesting and well-delineated definition of computation based on its behavioural capabilities.

One can see there are some strong parallelisms between this account and the concern of scale with what Floridi has names Levels of Abstraction \cite{floridi3} in that an information agent (the observer) accesses a physical or conceptual environment, the system, to determine whether it does compute or not. As with Levels of Abstraction, behavioural computational degrees as defined herein are not necessarily hierarchical but they are comparable and they may act as interfaces mediating between the epistemic relation of the observer and the observed.

In the physical world, under this qualitative approach, things may compute or not depending on the scale (or Level of Abstraction) at which they are studied. To say that a table computes only makes sense at the scale of the table, and as a $\mathbb{C}$-computer it should have a very limited $\mathbb{C}$, that is, a very limited behaviour given that it can hardly be programmed to do something else. Other possible objections are addressed in Ref.~\cite{sapere}.

\section{Concluding remarks}

I have proposed a novel qualitative notion of computation based on the sensitivity of a system to external stimuli connected to a concept of programmability, a notion I have called \emph{nature-like computation}, that provides a behavioural interpretation of computation (and of computers). This is consonant with current lines of technology for programming molecules and cells to compute. See for example Ref.~\cite{auslander}. In some sense this can be seen as reprogramming a cell to do certain tasks that it wasn't supposed to be able to do in the natural course of things. In a way this is what we have done with digital computers too, building machines out of natural matter to make them do calculations for us. Everything revolves around a single concept, that of programmability, which I have suggested can be captured by a measure of behaviour rather than by syntactic or even semantic approaches, given that the former requires descriptions of inner workings, even though we may not even fully understand the machinery of a cell, while the latter requires an interpretation of computation. The behavioural approach, however, is agnostic on most of these counts, being concerned only with the qualitative behaviour of a system, with its ability to transfer information upon being stimulated. The concept also helps to make sense of current uses of computation in the context of natural phenomena.

\section{Acknowledgments}

I am indebted to the generous reviewers whose comments have helped improve the presentation of this article. I also wish to thank the FQXi for the mini-grant awarded by way of the Silicon Valley Foundation under the title ``Time and Computation", in connection to behaviour as studied in this project (mini-grant no. 2011-93849 (4661)).

%REVISAR:
%Adami, Information Theory in Molecular Biology
%\url{http://arxiv.org/abs/q-bio/0405004}
%Putnam, H., 1967, ÒPsychological Predicates,Ó in Art, Mind, and Religion, W.H. Capitan and D.D. Merrill (eds.), Pittsburgh, PA: University of Pittsburgh Press, pp. 37Ð48. Reprinted in Putnam 1975a as ``The Nature of Mental States,Ó pp. 150Ð161.
%Putnam, H., 1975a, Philosophical Papers: Volume 2, Mind, Language and Reality, Cambridge: Cambridge University Press.
%ÐÐÐ, 1975b, ``The Meaning of `Meaning'," in Language, Mind and Knowledge. Minnesota Studies in the Philosophy of Science, vol. 7, K. Gunderson (ed.), Minneapolis: University of Minnesota Press, pp. 131Ð193. Reprinted in Putnam 1975a, pp. 215Ð271.
%Gualtiero Piccinini, Computation in Physical Systems, Stanford Encyclopedia of Philosophy, accessed on July 9, 2011.
%\url{http://www.ted.com/talks/george_dyson_at_the_birth_of_the_computer.html}
%buscar Putnam's theorem
%Dijkstra on philosophy of computation, programs and so on (programs make a system a computer...?)
%What is computation: http://www.cse.buffalo.edu/~rapaport/584/whatisanalg.html
%From http://serendip.brynmawr.edu/local/scisoc/information/albano22june04.html

\end{document}